\documentclass[11pt,twoside]{article}

\usepackage[latin1]{inputenc}
\usepackage{epsfig}
\usepackage{amsfonts}
\usepackage{cite}
\usepackage{color}
\definecolor{lgrey}{gray}{0.99}

\usepackage{graphicx}

\title{\huge
  Exact Solution of the Einstein Field Equations \\
  for a \\
  Spherical Shell of Fluid Matter}

\author{
  \Large Jorge L. deLyra\footnote{Email: delyra@lmail.if.usp.br} \\[2ex]
  \Large Rodrigo de A. Orselli\footnote{Email: orselli@if.usp.br} \\[2ex]
  \Large C. E. I. Carneiro\footnote{Email: ceugenio@if.usp.br} \\[2ex]
  Universidade de São Paulo \\
  Instituto de Física \\
  Rua do Matão, 1371, \\
  05508-090 São Paulo, SP, Brazil}

\date{April 5, 2021}

\newcommand{\FFrac}[2]{{\displaystyle\frac{\displaystyle #1}{\displaystyle #2}}}

\newcommand{\e}[1]{\,{\rm e}^{#1}}

\begin{document}\maketitle

\begin{abstract}
  \noindent
  We determine the exact solution of the Einstein field equations for the
  case of a spherically symmetric shell of liquid matter, characterized by
  an energy density which is constant with the Schwarzschild radial
  coordinate $r$ between two values $r_{1}$ and $r_{2}$. The solution is
  given in three regions, one being the well-known analytical
  Schwarzschild solution in the outer vacuum region, one being determined
  analytically in the inner vacuum region, and one being determined mostly
  analytically but partially numerically, within the matter region. The
  solutions for the temporal coefficient of the metric and for the
  pressure within this region are given in terms of a non-elementary but
  fairly straightforward real integral. For some values of the parameters
  this integral can be written in terms of elementary functions.
  
  We show that in this solution there is a singularity at the origin, and
  give the parameters of that singularity in terms of the geometrical and
  physical parameters of the shell. This does not correspond to an
  infinite concentration of matter, but in fact to zero energy density at
  the center. It does, however, imply that the spacetime within the
  spherical cavity is not flat, so that there is a non-trivial
  gravitational field there, in contrast with Newtonian gravitation. This
  gravitational field is repulsive with respect to the origin, and thus
  has the effect of stabilizing the geometrical configuration of the
  matter, since any particle of the matter that wanders out into either
  one of the vacuum regions tends to be brought back to the bulk of the
  matter by the gravitational field.
\end{abstract}

\newpage

\section{Introduction}\label{Sec01}

The Schwarzschild external solution~\cite{Schwarzschild,Wald} of the
Einstein field equations has played a major role in General Relativity. It
describes the effects of gravitation in the vacuum {\em outside} a
time-independent spherically symmetric distribution of matter. One of the
reasons for its importance is its generality --- it only depends on the
spherical symmetry and on the total energy of the matter distribution.
Jebsen and Birkhoff~\cite{JebsenTheorem,BirkhoffTheorem} have shown that
this solution is still valid even in time-dependent situations, provided
that the spherical symmetry is preserved. Another reason for its
popularity is the association of the coordinate singularity of this
solution, which occurs for a certain value of the radial coordinate, with
the presence of an event horizon, thus leading to the concept of black
holes.

Less known --- even absent in many standard textbooks on General
Relativity --- is the interior Schwarzschild
solution~\cite{SchwarzschildInternal,Wald}. It gives the metric of the
space {\em inside} a spherically symmetric matter distribution with an
energy density which is constant with the radial coordinate. This other
solution can be continuously joined with the Schwarzschild vacuum solution
that is valid outside the matter distribution. It is less general in that
it only describes matter distributions with energy densities that do not
depend on the radial coordinate $r$. In addition, it does not contain any
singularities. This point is emphasized in many texts, for example
in~\cite{Wald,MisnerThorneWheeler}. Basically, in order to avoid
singularities at the center of the matter distribution a certain
integration constant is set equal to zero.

For a spherical matter shell characterized by an inner radius $r_{1}$, an
outer radius $r_{2}$ and an energy density constant with $r$ the situation
is more involved. In the inner vacuum region, where $r<r_{1}$, the
solution of the Einstein equations leads to an integration constant,
heretofore denoted by $r_{\mu}$, which determines the singularities in the
entire inner vacuum region. There are no singularities only if
$r_{\mu}=0$. In analogy with what is done for the interior Schwarzschild
solution one may feel tempted to set $r_{\mu}=0$ by hand and eliminate all
singularities. However, as we are going to show in this paper, the correct
approach is to start in the outer vacuum region ($r>r_{2}$), where the
Schwarzschild external solution holds, and use the continuity of the
solution in the two boundaries of the three regions to determine the
constant $r_{\mu}$. The rather surprising result is that the imposition of
the surface boundary conditions implies that $r_{\mu}>0$, so that the
solutions do contain a singularity at the origin. In addition, one can
prove that this condition has to be satisfied in order to produce
solutions with non-negative pressure inside the matter shell.

It is remarkable that the boundary conditions on matter interfaces for the
Einstein field equations seem to play a smaller than expected role in the
literature. A rare example in which the role of these boundary conditions
is emphasized can be found in~\cite{XiaochuMei}, although the author of
that paper only obtained solutions containing a negative pressure region
inside the matter shell. By analyzing these negative pressure solutions
the author concluded that matter cannot collapse towards the center of
black holes in general relativity. We are going to show in this paper that
it is possible to obtain physically reasonable matter shell solutions of
the Einstein equations with non-negative and finite pressure inside the
shell. It is important to emphasize that the singularity at the origin in
the inner vacuum region does not lead to any divergence of the matter
quantities, and in fact stabilizes the matter shell structure. This is so
because the gravitational field within the inner vacuum region turns out
to be repulsive with respect to the origin. Our solutions for matter
shells are expressed in terms of a single integral which for some values
of the physical parameters can be written in terms of elementary functions
and constitute a new class of exact solutions of the Einstein field
equations.

Results similar to the ones we present here were obtained for the case of
neutron stars, with the Chandrasekhar equation of state~\cite{WeinbergGC},
by Ni~\cite{NiNeutrStars}, including the presence of inner and outer
matter-vacuum interfaces. However, the crucial consideration of the
interface boundary conditions was missing from that analysis, thus leading
to incomplete results. The discussion of the interface boundary conditions
was subsequently introduced by Neslu\v{s}an~\cite{NeslusanNeutrStars},
thus completing the analysis of the case of the neutron stars. Just as in
the present work, the discussion of the interface boundary conditions led,
also in that case, to an inner vacuum region containing a singularity at
the origin and a gravitational field pointing away from the origin, that
is, repulsive with respect to the origin. The present work can be
considered as an exactly solvable laboratory model that illustrates some
of the properties of that solution. It also shows that the properties of
the inner vacuum region are not artifacts of that particular problem or of
that particular type of equation of state.

This paper is organized as follows. In Section~\ref{Sec02} we state and
solve the problem; in Section~\ref{Sec03} we derive the main physical
properties of the solution; in Section~\ref{Sec04} we present a
two-parameter family of explicit solutions and a few numerical examples;
and in Section~\ref{Sec05} we present our conclusions.

\section{The Problem and its Solution}\label{Sec02}

We will present, in the case of a spherically symmetric shell of liquid
fluid with constant energy density, the exact solution of the Einstein
field equations of General Relativity~\cite{DiracGravity},

\begin{equation}\label{Eqn01}
  R_{\mu}^{\;\nu}-\frac{1}{2}\,R\,g_{\mu}^{\;\nu}
  =
  -\kappa\,T_{\mu}^{\;\nu},
\end{equation}

\noindent
where $\kappa=8\pi G/c^{4}$, $G$ is the universal gravitational constant
and $c$ is the speed of light. Under the conditions of time independence
and of spherical symmetry around the origin of a spherical system of
coordinates $(t,r,\theta,\phi)$, the Schwarzschild system of coordinates,
the most general possible metric is given by the invariant interval,
written in terms of this spherical system of coordinates,

\begin{equation}\label{Eqn02}
  ds^{2}
  =
  \e{2\nu(r)}c^{2}dt^{2}
  -
  \e{2\lambda(r)}dr^{2}
  -
  r^{2}\left[d\theta^{2}+\sin^{2}(\theta)d\phi^{2}\right],
\end{equation}

\noindent
where $\exp[\nu(r)]$ and $\exp[\lambda(r)]$ are two positive functions of
only $r$. As one can see, in this work we will use the time-like signature
$(+,-,-,-)$, following~\cite{DiracGravity}. Under these conditions the
matter energy-momentum tensor density $T_{\mu}^{\;\nu}$ on the right-hand
side of the equation is diagonal, and given by the four diagonal
components $T_{0}^{\;0}(r)=\rho(r)$, where $\rho(r)$ is the energy density
of the matter, and $T_{1}^{\;1}(r)=T_{2}^{\;2}(r)=T_{3}^{\;3}(r)=-P(r)$,
where $P(r)$ is the pressure, which is isotropic, thus characterizing a
fluid.

Since under these conditions $R_{\mu}^{\;\nu}$ and $T_{\mu}^{\;\nu}$ are
both diagonal, there are just four non-trivial field equations contained
in Equation~(\ref{Eqn01}). In addition to these four field equations we
have the consistency condition

\begin{equation}\label{Eqn03}
  D_{\nu}T_{\mu}^{\;\nu}
  =
  0,
\end{equation}

\noindent
which is due to the fact that the combination of tensors that constitutes
the left-hand side of the Einstein field equation satisfies the Bianchi
identity of the Ricci curvature tensor. Writing these equations explicitly
in the chosen coordinate system, one finds that the component equations
involving $T_{2}^{\;2}(r)$ and $T_{3}^{\;3}(r)$ turn out to be identical,
so that we are left with the set of four equations, including the
consistency condition,

\noindent
\begin{eqnarray}
  \label{Eqn04}
  \left\{\rule{0em}{3ex}1-2\left[r\lambda'(r)\right]\right\}
  \e{-2\lambda(r)}
  & = &
        1-\kappa r^{2}\rho(r),
  \\
  \label{Eqn05}
  \left\{\rule{0em}{3ex}1+2\left[r\nu'(r)\right]\right\}
  \e{-2\lambda(r)}
  & = &
        1+\kappa r^{2}P(r),
  \\
  \label{Eqn06}
  \left\{\rule{0em}{3ex}
  r^{2}\nu''(r)
  -
  \left[r\lambda'(r)\right]\left[r\nu'(r)\right]
  \right.
  \hspace{7em}
  &   &
        \nonumber
  \\
  \left.\rule{0em}{3ex}
  +
  \left[r\nu'(r)\right]^{2}
  +
  \left[r\nu'(r)\right]
  -
  \left[r\lambda'(r)\right]
  \right\}
  \e{-2\lambda(r)}
  & = &
        \kappa r^{2}P(r),
  \\
  \label{Eqn07}
  \left[\rho(r)+P(r)\right]
  \nu'(r)
  & = &
        -P'(r),
\end{eqnarray}

\noindent
where the primes indicate differentiation with respect to $r$. Next, it
can be shown that Equation~(\ref{Eqn06}) can be obtained from the others,
being in fact a linear combination of the derivative of
Equation~(\ref{Eqn05}) and of Equations~(\ref{Eqn04}),~(\ref{Eqn05})
and~(\ref{Eqn07}). If we denote Equations~(\ref{Eqn04})
through~(\ref{Eqn07}) respectively by $E_{t}$, $E_{r}$, $E_{\theta}$ and
$E_{c}$, we have that

\begin{equation}\label{Eqn08}
  E_{\theta}
  =
  \frac{1}{2}
  \left[
    -r\nu'(r)\left(E_{t}-E_{r}\right)
    +
    rE'_{r}
    +
    \kappa r^{2}E_{c}
  \right].
\end{equation}

\noindent
This leaves us with a set of just three differential equations to
solve. In addition to this, we will assume that we have an energy density
$\rho(r)=\rho_{0}$ which is constant as a function of $r$ within the shell
of fluid matter, thus characterizing a liquid fluid. The equations that we
propose to solve are therefore those given in
Equations~(\ref{Eqn04}),~(\ref{Eqn05}) and~(\ref{Eqn07}). It is important
to note that, in this way, we are left with a system of just three {\em
  first-order} differential equations. Therefore, the discussion of
boundary conditions can be limited to the discussion of the behavior of
the functions involved, thus eliminating the need for any discussion of
the behavior of their derivatives.

We will assume that the matter consists of a spherical shell of liquid,
located between the radial positions $r_{1}$ and $r_{2}$, meaning that we
will have an inner vacuum region within $(0,r_{1})$, a matter region
within $(r_{1},r_{2})$, and an outer vacuum region within
$(r_{2},\infty)$. This means that we will have for $\rho(r)$ and $P(r)$

\noindent
\begin{eqnarray}\label{Eqn09}
  \rho(r)
  & = &
        \left\{
        \begin{array}{lcl}
          0
          &
            \mbox{for}
          &
            0\;\leq r<r_{1},
          \\[3ex]
          \rho_{0}
          &
            \mbox{for}
          &
            r_{1}<r<r_{2},
          \\[3ex]
          0
          &
            \mbox{for}
          &
            r_{2}<r<\infty,
        \end{array}
            \right.
            \nonumber
  \\[3ex]
  P(r)
  & = &
        \left\{
        \begin{array}{lcl}
          0
          &
            \mbox{for}
          &
            0\;\leq r\leq r_{1},
          \\[3ex]
          0
          &
            \mbox{for}
          &
            r_{2}\leq r<\infty.
        \end{array}
            \right.
\end{eqnarray}

\noindent
The function $P(r)$ within the matter region is, of course, one of the
unknowns of our problem. In addition to this, we have the boundary
conditions for $P(r)$ at the two interfaces, in the limits coming from
within the liquid,

\noindent
\begin{eqnarray}\label{Eqn10}
  P(r_{1})
  & = &
        0,
        \nonumber
  \\
  P(r_{2})
  & = &
        0,
\end{eqnarray}

\noindent
since these constitute a requirement in any interface between fluid matter
and a vacuum. The remaining boundary conditions are those requiring the
continuity of $\lambda(r)$ and $\nu(r)$ across the interfaces, and the
asymptotic conditions leading to the Newtonian limit at radial infinity.

\subsection{Solutions in the Vacuum Regions}\label{SSec0202}

Within either vacuum region the consistency condition in
Equation~(\ref{Eqn07}) becomes a mere identity, so that we are left with
only two equations, in which we replace both $\rho(r)$ and $P(r)$ by zero,

\noindent
\begin{eqnarray}\label{Eqn11}
  1-2\left[r\lambda'(r)\right]
  & = &
        \e{2\lambda(r)},
        \nonumber
  \\
  1+2\left[r\nu'(r)\right]
  & = &
        \e{2\lambda(r)}.
\end{eqnarray}

\noindent
This immediately implies that $\lambda'(r)+\nu'(r)=0$, and hence that
$\lambda(r)+\nu(r)=A$, where $A$ is a dimensionless integration
constant. The first of these two equations involves only $\lambda(r)$, and
can also be written as

\begin{equation}\label{Eqn12}
  \left[r\e{-2\lambda(r)}\right]'
  =
  1,
\end{equation}

\noindent
which can be immediately integrated to

\begin{equation}\label{Eqn13}
  \e{-2\lambda(r)}
  =
  1-\frac{R}{r},
\end{equation}

\noindent
where $R$ is an integration constant with dimensions of length.

We must now discriminate between the inner and outer vacuum regions. In
the outer vacuum region we must get flat space at radial infinity, which
requires that both $\lambda(r)$ and $\nu(r)$ go to zero for
$r\to\infty$. This in turn implies that $A=0$ in the outer vacuum region,
thus leading to $\nu(r)=-\lambda(r)$. As is well known, the condition that
the Newtonian limit be realized at radial infinity requires that
$R=r_{M}$, the Schwarzschild radius $r_{M}=2MG/c^{2}$ associated to the
asymptotic gravitational mass $M$ of the system. Thus we arrive at the
time-honored Schwarzschild solution~\cite{Schwarzschild,Wald} in the outer
vacuum region,

\noindent
\begin{eqnarray}\label{Eqn14}
  \lambda_{s}(r)
  & = &
        -\,
        \frac{1}{2}\,
        \ln\!\left(\frac{r-r_{M}}{r}\right),
        \nonumber
  \\
  \nu_{s}(r)
  & = &
        \frac{1}{2}\,
        \ln\!\left(\frac{r-r_{M}}{r}\right),
\end{eqnarray}

\noindent
where the subscript $s$ denotes the outer vacuum region. Note that there
is a limitation on the values of the parameters $r_{2}$ and $r_{M}$
describing the distribution of matter, because these expressions have a
singular behavior at $r=r_{M}$. We must have $r_{M}<r_{2}$ to ensure that
there is no event horizon formed outside the distribution of matter.

In the inner vacuum region there are no asymptotic conditions to be
applied, and thus the integration constants $A$ and $R$ will have to be
left undetermined, to be determined later on via the boundary conditions
at the interfaces between the vacuum and the matter, as we come in from
radial infinity towards the origin. For convenience we will put
$R=-r_{\mu}$, and write the solution in the inner vacuum region as

\noindent
\begin{eqnarray}\label{Eqn15}
  \lambda_{i}(r)
  & = &
        -\,
        \frac{1}{2}\,
        \ln\!\left(\frac{r+r_{\mu}}{r}\right),
        \nonumber
  \\
  \nu_{i}(r)
  & = &
        A
        +
        \frac{1}{2}\,
        \ln\!\left(\frac{r+r_{\mu}}{r}\right),
\end{eqnarray}

\noindent
where the subscript $i$ denotes the inner vacuum region. Note that the
value of $r_{\mu}$ determines the singularity structure of this solution
within the inner vacuum region. If $r_{\mu}<0$ then there is a singularity
at the strictly positive radial position $r=-r_{\mu}$, corresponding to
the formation of an event horizon at that position. If $r_{\mu}=0$ then
there are no singularities at all within this region. If $r_{\mu}>0$ then
there is only one singularity, located at the origin $r=0$. We will show
later on that we do indeed have that $r_{\mu}>0$.

We therefore have the complete analytical solutions in the inner and outer
vacuum regions, which contain one input parameter of the problem, the mass
$M$ associated to the Schwarzschild radius $r_{M}$, and two integration
constants still to be determined, $A$ and $r_{\mu}$.

\subsection{Solution in the Matter Region}\label{SSec0203}

In the matter region Equation~(\ref{Eqn04}) for $\lambda(r)$ can be
written as

\begin{equation}\label{Eqn16}
  \left[r\e{-2\lambda(r)}\right]'
  =
  1-\kappa\rho_{0} r^{2},
\end{equation}

\noindent
which can be immediately integrated to

\begin{equation}\label{Eqn17}
  \e{-2\lambda(r)}
  =
  1
  +
  \frac{B}{r}
  -
  \frac{\kappa\rho_{0}}{3}\,r^{2},
\end{equation}

\noindent
where $B$ is an integration constant with dimensions of length, thus
leading to the general solution for $\lambda(r)$ in the matter region,

\begin{equation}\label{Eqn18}
  \lambda_{m}(r)
  =
  -\,
  \frac{1}{2}\,
  \ln\!\left(1+\frac{B}{r}-\frac{\kappa\rho_{0}}{3}\,r^{2}\right),
\end{equation}

\noindent
where the subscript $m$ denotes the matter region. This solution contains
one integration constant, the constant $B$, and one parameter
characterizing the system, namely $\rho_{0}$, which is not, however, a
free input parameter of the problem, since it will depend on $M$ and thus
on $r_{M}$.

In order to deal with $\nu(r)$ in the matter region, we consider the
consistency condition given in Equation~(\ref{Eqn07}), which can be
written in this region as

\begin{equation}\label{Eqn19}
  \nu'(r)
  =
  -\,
  \frac{P'(r)}{\rho_{0}+P(r)},
\end{equation}

\noindent
thus allowing us to separate variables and hence to write $\nu(r)$ in
terms of $P(r)$,

\noindent
\begin{eqnarray}\label{Eqn20}
  d\nu
  & = &
        -\,
        \frac{dP}{\rho_{0}+P}
        \nonumber
  \\
  & = &
        -d\ln\!\left(\rho_{0}+P\right).
\end{eqnarray}

\noindent
If we integrate from the left end $r_{1}$ of the matter interval to a
generic point $r$ within that interval, we get

\begin{equation}\label{Eqn21}
  \nu(r)-\nu(r_{1})
  =
  -
  \ln\!\left[\frac{\rho_{0}+P(r)}{\rho_{0}+P(r_{1})}\right].
\end{equation}

\noindent
However, the boundary conditions for $P(r)$ at the interfaces tell us that
we must have $P(r_{1})=0$, and hence we get the general solution for
$\nu(r)$ within the matter region, written in terms of $P(r)$,

\begin{equation}\label{Eqn22}
  \nu_{m}(r)
  =
  \nu_{1}
  -
  \ln\!\left[\frac{\rho_{0}+P(r)}{\rho_{0}}\right],
\end{equation}

\noindent
where $\nu_{1}=\nu(r_{1})$. The solutions for $\lambda(r)$ and $\nu(r)$
within the matter region involve therefore two integration constants, $B$
and $\nu_{1}$. The solution for $\nu(r)$ is not yet completely determined,
since it is given in terms of $P(r)$, which is also as yet
undetermined. However, the information obtained so far already allows us
to impose the boundary conditions at the interfaces, in order to determine
the integration constants, which is what we turn to now.

\subsection{Interface Boundary Conditions}\label{SSec0204}

The condition of the continuity of $\lambda(r)$ at the interface $r_{1}$
implies that we must have that $\lambda_{i}(r_{1})=\lambda_{m}(r_{1})$,
which from Equations~(\ref{Eqn15}) and~(\ref{Eqn18}) gives us the
following relation between the parameters

\begin{equation}\label{Eqn23}
  B-r_{\mu}
  =
  \frac{\kappa\rho_{0}}{3}\,r_{1}^{3}.
\end{equation}

\noindent
In addition to this, the condition of the continuity of $\lambda(r)$ at
the interface $r_{2}$ implies that we must have
$\lambda_{m}(r_{2})=\lambda_{s}(r_{2})$, which from
Equations~(\ref{Eqn14}) and~(\ref{Eqn18}) gives us the following relation
between the parameters

\begin{equation}\label{Eqn24}
  B+r_{M}
  =
  \frac{\kappa\rho_{0}}{3}\,r_{2}^{3}.
\end{equation}

\noindent
This last condition already determines the integration constant $B$ in
terms of the parameters of the problem,

\begin{equation}\label{Eqn25}
  B
  =
  -
  r_{M}
  +
  \frac{\kappa\rho_{0}}{3}\,r_{2}^{3},
\end{equation}

\noindent
and the difference of the two conditions just obtained determines the
integration parameter $r_{\mu}$ in terms of the parameters of the problem,

\begin{equation}\label{Eqn26}
  r_{\mu}
  =
  -
  r_{M}
  +
  \frac{\kappa\rho_{0}}{3}\left(r_{2}^{3}-r_{1}^{3}\right).
\end{equation}

\noindent
We have therefore the solution for $\lambda(r)$ in the matter region, in
terms of the parameters of the problem,

\begin{equation}\label{Eqn27}
  \lambda_{m}(r)
  =
  -\,
  \frac{1}{2}\,
  \ln\!
  \left[
    \frac
    {\kappa\rho_{0}\left(r_{2}^{3}-r^{3}\right)+3\left(r-r_{M}\right)}
    {3r}
  \right].
\end{equation}

\noindent
Let us point out that there is a consistency condition to be applied to
this result, since we must have that the cubic polynomial appearing in the
argument of the logarithm be strictly positive for all values of $r$
within the matter region, that is

\begin{equation}\label{Eqn28}
  \kappa\rho_{0}
  \left(r_{2}^{3}-r^{3}\right)+3\left(r-r_{M}\right)
  >
  0,
\end{equation}

\noindent
for all $r\in[r_{1},r_{2}]$. Note that the term with the cubes is
necessarily non-negative, but that the other term may be negative, if
$r_{M}$ is not smaller than $r_{1}$. Therefore, so long as $r_{M}<r_{1}$,
this strict positivity condition is automatically satisfied. If, however,
we have that $r_{1}<r_{M}<r_{2}$, then the condition must be actively
verified for all $r\in[r_{M},r_{2}]$. If it fails, then there is no
solution for that particular set of input parameters.

Since we have $\nu_{m}(r)$ written in terms of $P(r)$, and since we know
the interface boundary conditions for $P(r)$ in limits from within the
matter region, we are in a position to impose the boundary conditions on
$\nu(r)$ across the interfaces, even without having available the complete
solution for $\nu_{m}(r)$. To this end, let us note that from
Equation~(\ref{Eqn22}) we have that
$\nu_{m}(r_{1})=\nu_{m}(r_{2})=\nu_{1}$. At the interface $r_{1}$ the
condition of the continuity of $\nu(r)$ implies that we must have
$\nu_{i}(r_{1})=\nu_{m}(r_{1})$, which from Equations~(\ref{Eqn15})
and~(\ref{Eqn22}) gives us the following relation between the parameters,

\begin{equation}\label{Eqn29}
  \nu_{1}
  =
  A
  +
  \frac{1}{2}\,
  \ln\!\left(\frac{r_{1}+r_{\mu}}{r_{1}}\right).
\end{equation}

\noindent
In addition to this, the condition of the continuity of $\nu(r)$ at the
interface $r_{2}$ implies that we must have
$\nu_{m}(r_{2})=\nu_{s}(r_{2})$, which from Equations~(\ref{Eqn14})
and~(\ref{Eqn22}) gives us the following relation between the parameters,

\begin{equation}\label{Eqn30}
  \nu_{1}
  =        
  \frac{1}{2}\,
  \ln\!\left(\frac{r_{2}-r_{M}}{r_{2}}\right).
\end{equation}

\noindent
This last condition gives us the integration constant $\nu_{1}$ in terms
of the parameters of the problem, and its difference with the previous one
determines the integration constant $A$,

\begin{equation}\label{Eqn31}
  A
  =
  \frac{1}{2}\,
  \ln\!
  \left(
    \frac{r_{1}}{r_{2}}\,
    \frac{r_{2}-r_{M}}{r_{1}+r_{\mu}}
  \right).
\end{equation}

\noindent
This completes the determination of the solution for both $\nu(r)$ and
$\lambda(r)$ in the inner vacuum region, for which we now have

\noindent
\begin{eqnarray}\label{Eqn32}
  \lambda_{i}(r)
  & = &
        -\,
        \frac{1}{2}\,
        \ln\!\left(\frac{r+r_{\mu}}{r}\right),
        \nonumber
  \\
  \nu_{i}(r)
  & = &
        \frac{1}{2}\,
        \ln\!
        \left(
        \frac{r_{1}}{r_{2}}\,
        \frac{r_{2}-r_{M}}{r_{1}+r_{\mu}}
        \right)
        +
        \frac{1}{2}\,
        \ln\!\left(\frac{r+r_{\mu}}{r}\right),
\end{eqnarray}

\noindent
with $r_{\mu}$ given by Equation~(\ref{Eqn26}). We also have the following
form for the solution for $\nu(r)$ within the matter region, still in
terms of $P(r)$,

\begin{equation}\label{Eqn33}
  \nu_{m}(r)
  =
  \frac{1}{2}\,
  \ln\!\left(\frac{r_{2}-r_{M}}{r_{2}}\right)
  -
  \ln\!\left[\frac{\rho_{0}+P(r)}{\rho_{0}}\right].
\end{equation}

\noindent
At this point the situation is as follows, in regard to the complete
solution of the problem. Given values of $r_{1}$, $r_{2}$ and $r_{M}$,
which completely characterize the geometrical and physical nature of the
object under study, we have the complete solution for both $\lambda(r)$
and $\nu(r)$ in the outer vacuum region. We also have the complete
solution for both $\lambda(r)$ and $\nu(r)$ in the inner vacuum region,
except for the determination of the parameter $\rho_{0}$. We have as well
the complete solution for $\lambda(r)$ in the matter region, again up to
the determination of the parameter $\rho_{0}$. The one element of the
solution still missing is the complete solution for $\nu(r)$ in the matter
region. However, since we have $\nu(r)$ determined in terms of $P(r)$ in
this region, this can also be accomplished by the complete determination
of $P(r)$ in this region, which is the task we tackle next. Let us
emphasize that the parameter $\rho_{0}$ is not a free input parameter of
the problem, since it must be chosen so that the given value of $r_{M}$
results, that is, the local value of the energy density must be chosen so
that the given value of the asymptotic gravitational mass $M$ results at
radial infinity.

\subsection{The Equation for the Pressure}\label{SSec0205}

The equation determining the pressure $P(r)$ in the matter region can be
obtained by eliminating $\nu'(r)$ from Equations~(\ref{Eqn05})
and~(\ref{Eqn07}), which gives us

\begin{equation}\label{Eqn34}
  \rho_{0}+P(r)-2\left[rP'(r)\right]
  =
  \e{2\lambda_{m}(r)}
  \left[1+\kappa r^{2}P(r)\right]\left[\rho_{0}+P(r)\right].
\end{equation}

\noindent
In this equation the quantity $\exp[2\lambda_{m}(r)]$ is a known function,
since we have already determined $\lambda(r)$ in the matter region. This
is a first-order non-linear differential equation determining $P(r)$, with
the boundary conditions $P(r_{1})=0$ and $P(r_{2})=0$. Since the equation
is first-order and there are two boundary conditions to be satisfied, it
is clear that the parameter $\rho_{0}$ will have to be adjusted so that
the second condition can be satisfied. This will therefore determine the
parameter $\rho_{0}$ in terms of the other parameters of the problem. This
equation can be simplified by a series of transformations on the variables
and parameters. First we define the parameter $\Upsilon_{0}$, which has
dimensions of inverse length and is such that

\begin{equation}\label{Eqn35}
  \Upsilon_{0}^{2}
  =
  \kappa\rho_{0},
\end{equation}

\noindent
and the dimensionless pressure

\begin{equation}\label{Eqn36}
  p(r)
  =
  \frac{P(r)}{\rho_{0}},
\end{equation}

\noindent
in terms of which Equation~(\ref{Eqn34}) becomes

\begin{equation}\label{Eqn37}
  \left[rp'(r)\right]
  =
  \frac{1}{2}\left[1+p(r)\right]
  \left\{
    1-\e{2\lambda_{m}(r)}\left[1+\Upsilon_{0}^{2}r^{2}p(r)\right]
  \right\}.
\end{equation}

\noindent
Substituting the known value of $\lambda_{m}(r)$ from
Equation~(\ref{Eqn27}) we get

\begin{equation}\label{Eqn38}
  p'(r)
  =
  \frac{1}{2r}\left[1+p(r)\right]
  \frac
  {
    \Upsilon_{0}^{2}\left(r_{2}^{3}-r^{3}\right)
    -
    3r_{M}
    -
    3\Upsilon_{0}^{2}r^{3}p(r)
  }
  {
    \Upsilon_{0}^{2}
    \left(r_{2}^{3}-r^{3}\right)+3\left(r-r_{M}\right)
  }.
\end{equation}

\noindent
This has the form of a Riccati equation, and can be linearized by the
transformation of variables

\begin{equation}\label{Eqn39}
  p(r)
  =
  \frac{1}{z(r)}-1,
\end{equation}

\noindent
thus resulting in the equation for $z(r)$,

\begin{equation}\label{Eqn40}
  z'(r)
  +
  \frac
  {\Upsilon_{0}^{2}\left(r_{2}^{3}+2r^{3}\right)-3r_{M}}
  {
    2r
    \left[
      \Upsilon_{0}^{2}\left(r_{2}^{3}-r^{3}\right)
      +
      3\left(r-r_{M}\right)
    \right]
  }\,
  z(r)
  =
  \frac
  {3\Upsilon_{0}^{2}r^{3}}
  {
    2r
    \left[
      \Upsilon_{0}^{2}\left(r_{2}^{3}-r^{3}\right)
      +
      3\left(r-r_{M}\right)
    \right]
  }.
\end{equation}

\noindent
This equation has an integrating factor given by $\exp[F(r)]$, where
$F(r)$ is defined as an integral of the coefficient of the second term
from $r_{2}$ to some arbitrary $r$ within $[r_{1},r_{2}]$,

\noindent
\begin{eqnarray}\label{Eqn41}
  F(r)
  & = &
        \int_{r_{2}}^{r}ds\,
        \frac
        {\Upsilon_{0}^{2}\left(r_{2}^{3}+2s^{3}\right)-3r_{M}}
        {
        2s
        \left[
        \Upsilon_{0}^{2}\left(r_{2}^{3}-s^{3}\right)
        +
        3\left(s-r_{M}\right)
        \right]
        }
        \nonumber
  \\
  & = &
        \frac{1}{2}
        \int_{r_{2}}^{r}ds\,
        \frac{1}{s}
        -
        \frac{1}{2}
        \int_{r_{2}}^{r}ds\,
        \frac
        {-3\Upsilon_{0}^{2}s^{2}+3}
        {
        \Upsilon_{0}^{2}
        \left(r_{2}^{3}-s^{3}\right)+3\left(s-r_{M}\right)
        }.
\end{eqnarray}

\noindent
One can see now that both integrals can be done, and thus we obtain

\begin{equation}\label{Eqn42}
  e^{F(r)}
  =
  \sqrt{\frac{r}{r_{2}}}
  \;
  \sqrt
  {
    \frac
    {3\left(r_{2}-r_{M}\right)}
    {
      \Upsilon_{0}^{2}
      \left(r_{2}^{3}-r^{3}\right)+3\left(r-r_{M}\right)
    }
  },
\end{equation}

\noindent
in terms of which the equation for $z(r)$ can be written as

\begin{equation}\label{Eqn43}
  \left[\e{F(r)}z(r)\right]'
  =
  \frac{3}{2}\,
  \frac
  {\Upsilon_{0}^{2}r^{2}\e{F(r)}}
  {\Upsilon_{0}^{2}\left(r_{2}^{3}-r^{3}\right)+3\left(r-r_{M}\right)},
\end{equation}

\noindent
which can then be integrated over the interval $[r_{2},r]$ giving

\begin{equation}\label{Eqn44}
  z(r)
  =
  e^{-F(r)}+\frac{3}{2}\,e^{-F(r)}\!
  \int_{r_{2}}^{r}ds\,
  \frac
  {\Upsilon_{0}^{2}s^{2}e^{F(s)}}
  {\Upsilon_{0}^{2}\left(r_{2}^{3}-s^{3}\right)+3\left(s-r_{M}\right)},
\end{equation}

\noindent
where we used the fact that by definition $F(r_{2})=0$, and the fact that
$P(r_{2})=0$ implies $z(r_{2})=1$.

Note that once more the existence of the solutions for $F(r)$ and for
$z(r)$ is conditioned by the strict positivity of the same cubic
polynomial that we discussed before in Equation~(\ref{Eqn28}), which we
can now write as

\begin{equation}\label{Eqn45}
  \Upsilon_{0}^{2}\left(r_{2}^{3}-r^{3}\right)
  +
  3\left(r-r_{M}\right)
  >
  0,
\end{equation}

\noindent
for all $r\in[r_{1},r_{2}]$. Substituting the value of $\exp[F(r)]$ we
have the solution for $z(r)$ written in terms of a real integral,

\noindent
\begin{eqnarray}\label{Eqn46}
  z(r)
  & = &
        \sqrt
        {
        \frac
        {
        \Upsilon_{0}^{2}\left(r_{2}^{3}-r^{3}\right)
        +
        3\left(r-r_{M}\right)
        }
        {r}
        }
        \nonumber
  \\
  &   &
        \times
        \left\{
        \sqrt{\frac{r_{2}}{3\left(r_{2}-r_{M}\right)}}
        +
        \frac{3}{2}\,
        \int_{r_{2}}^{r}ds\,
        \frac
        {\Upsilon_{0}^{2}s^{5/2}}
        {
        \left[
        \Upsilon_{0}^{2}\left(r_{2}^{3}-s^{3}\right)
        +
        3\left(s-r_{M}\right)
        \right]^{3/2}
        }
        \right\}.
\end{eqnarray}

\noindent
In most cases this remaining integral is elliptic and therefore cannot be
written in terms of elementary functions, so that in general this
remaining last step of the resolution procedure has to be performed by
numerical means. However, as we are going to show in Section~\ref{Sec04},
for some values of the parameters it is possible to express this integral
in terms of elementary functions.

After determining $z(r)$ in the matter region, Equations~(\ref{Eqn39})
allows us to calculate the dimensionless pressure $p(r)$ which, according
to Equation~(\ref{Eqn36}), is equal to the pressure divided by the energy
density $\rho_{0}$,

\noindent
\begin{eqnarray}\label{Eqn47}
  p(r)
  & = &
        \frac{1}{z(r)}-1
        \;\;\;\Longrightarrow
        \nonumber
  \\
  P(r)
  & = &
        \frac{\rho_{0}}{z(r)}-\rho_{0}.
\end{eqnarray}

\noindent
Note that $z(r)$ also determines $\nu(r)$ in the matter region, since in
Equation~(\ref{Eqn33}) we have $\nu_{m}(r)$ in terms of $P(r)$, and
therefore we have for the exponential of $\nu_{m}(r)$,

\begin{equation}\label{Eqn48}
  \e{\nu_{m}(r)}
  =
  \sqrt{\frac{r_{2}-r_{M}}{r_{2}}}\,
  \frac{\rho_{0}}{\rho_{0}+P(r)},
\end{equation}

\noindent
which, using Equation~(\ref{Eqn47}), implies that

\begin{equation}\label{Eqn49}
  \e{\nu_{m}(r)}
  =
  \sqrt{\frac{r_{2}-r_{M}}{r_{2}}}\,
  z(r),
\end{equation}

\noindent
so that, up to a constant factor, $z(r)$ turns out to be the square root
of the temporal coefficient of the metric. This completes the
determination of the solution in all three regions, in terms of the
parameters of the problem. Given certain values of $r_{1}$, $r_{2}$ and
$r_{M}$, one must still find a value of the parameter $\rho_{0}$, and
hence of $\Upsilon_{0}$, such that the boundary conditions for $P(r)$ at
the two interfaces are satisfied. One can obtain an equation determining
this value of $\Upsilon_{0}$ by considering the value of $z(r_{1})$. Since
$P(r_{1})=0$, we have that $z(r_{1})=1$, so that from
Equation~(\ref{Eqn46}) we get

\noindent
\begin{eqnarray}\label{Eqn50}
  \sqrt{\frac{r_{2}}{3\left(r_{2}-r_{M}\right)}}
  & = &
        \sqrt
        {
        \frac{r_{1}}
        {
        \Upsilon_{0}^{2}\left(r_{2}^{3}-r_{1}^{3}\right)
        +
        3\left(r_{1}-r_{M}\right)
        }
        }
        \nonumber
  \\
  &   &
        +
        \frac{3}{2}
        \int_{r_{1}}^{r_{2}}dr\,
        \frac
        {\Upsilon_{0}^{2}r^{5/2}}
        {
        \left[
        \Upsilon_{0}^{2}\left(r_{2}^{3}-r^{3}\right)
        +
        3\left(r-r_{M}\right)
        \right]^{3/2}
        }.
\end{eqnarray}

\noindent
The solution of this algebraic equation gives the value of $\Upsilon_{0}$,
and hence the value of $\rho_{0}$, for which the two interface boundary
conditions for $P(r)$ will be satisfied. The solution of this equation
necessarily includes the consistency check of the solution obtained, since
the calculation of the integral is dependent on the strict positivity of
the polynomial in Equation~(\ref{Eqn45}), for all $r$ within
$[r_{1},r_{2}]$. This is the same condition that guarantees the
consistency of the results for $F(r)$ and $z(r)$, and hence the
consistency of the results for $P(r)$ and $\nu(r)$ within the matter
region.

\section{Main Properties of the Solution}\label{Sec03}

In this section we will state and prove a few important properties of the
solution. We will assume that, given certain values of $r_{1}$, $r_{2}$
and $r_{M}$, the corresponding solution exists. In other words, we are
assuming that a solution of Equation~(\ref{Eqn50}) for $\Upsilon_{0}$ can
be found, thus determining $\rho_{0}$, which includes establishing the
strict positivity of the cubic polynomial within the square roots in the
denominators, and that a corresponding function $z(r)$ is therefore
determined via Equation~(\ref{Eqn46}). This then implies that the
solutions for both $\lambda(r)$ and $\nu(r)$, as well as for $P(r)$, are
all determined, with all the boundary conditions duly satisfied. A simpler
way to put this is to say that we are establishing the most important
properties of all existing solutions of the problem. For easy reference,
we state the complete solution explicitly in Table~\ref{Tab01}, where we
have that $\rho_{0}$ is determined algebraically via
Equation~(\ref{Eqn50}), $z(r)$ is determined by Equation~(\ref{Eqn46}),
and $r_{\mu}$ is given by Equation~(\ref{Eqn26}). We will start by the
discussion of the presence of the singularity at the origin.

\begin{table}
  \centering
  \caption{Summary of the results.}\label{Tab01}
  \fbox {
    \begin{minipage}[c]{32em}
      \begin{eqnarray*}
        \lambda(r)
        & = &
              \left\{
              \begin{array}{lcl}
                -\,
                \FFrac{1}{2}\,
                \ln\!\left(\FFrac{r+r_{\mu}}{r}\right)
                &
                  \mbox{for}
                &
                  0\;\leq r\leq r_{1},
                \\[3ex]
                -\,
                \FFrac{1}{2}\,
                \ln\!
                \left[
                \FFrac
                {
                \kappa\rho_{0}
                \left(r_{2}^{3}-r^{3}\right)
                +
                3\left(r-r_{M}\right)
                }
                {
                3r
                }
                \right]
                &
                  \mbox{for}
                &
                  r_{1}\leq r\leq r_{2},
                \\[3ex]
                -\,
                \FFrac{1}{2}\,
                \ln\!\left(\FFrac{r-r_{M}}{r}\right)
                &
                  \mbox{for}
                &
                  r_{2}\leq r<\infty,
              \end{array}
                  \right.
      \end{eqnarray*}
      \begin{eqnarray*}
        \nu(r)
        & = &
              \left\{
              \begin{array}{lcl}
                \FFrac{1}{2}\,
                \ln\!
                \left(
                \FFrac{r_{1}}{r_{2}}\,
                \FFrac{r_{2}-r_{M}}{r_{1}+r_{\mu}}
                \right)
                +
                \FFrac{1}{2}\,
                \ln\!\left(\FFrac{r+r_{\mu}}{r}\right)
                &
                  \mbox{for}
                &
                  0\;\leq r\leq r_{1},
                \\[3ex]
                \FFrac{1}{2}\,
                \ln\!
                \left(
                \FFrac{r_{2}-r_{M}}{r_{2}}
                \right)
                +
                \ln\!\left[z(r)\right]
                &
                  \mbox{for}
                &
                  r_{1}\leq r\leq r_{2},
                \\[3ex]
                \FFrac{1}{2}\,
                \ln\!\left(\FFrac{r-r_{M}}{r}\right)
                &
                  \mbox{for}
                &
                  r_{2}\leq r<\infty.
              \end{array}
                  \right.
      \end{eqnarray*}
      \\[-3.5ex]
    \end{minipage}
  }
\end{table}

\subsection{Existence of the Singularity at the Origin}\label{SSec0301}

The existence of the singularity at the origin is equivalent to the
statement that $r_{\mu}>0$, because the only way to avoid that singularity
would be to have $r_{\mu}=0$. If we put $r_{\mu}=0$ and take the limit
$r_{1}\rightarrow 0$ we no longer have a matter shell, and we obtain
instead the Schwarzschild interior solution.

We start with a preliminary lemma, in which we will prove that the
following combination of parameters

\begin{equation}\label{Eqn51}
  \frac{1}{3}\,
  \Upsilon_{0}^{2}\left(r_{2}^{3}-r_{e}^{3}\right)
  -
  r_{M}
  >
  0,
\end{equation}

\noindent
is strictly positive, where $r_{e}$ is the position of the maximum of the
dimensionless pressure $p(r)$ within the interval $[r_{1},r_{2}]$. In
order to do this, we consider the equation for $p(r)$ given in
Equation~(\ref{Eqn38}). Applying that equation at $r_{2}$, since we have
that $p(r_{2})=0$, we get for the derivative at the right end of the
matter interval,

\begin{equation}\label{Eqn52}
  p'(r_{2})
  =
  -\,
  \frac{r_{M}}{2r_{2}\left(r_{2}-r_{M}\right)}.
\end{equation}

\noindent
Since by hypothesis we have that $r_{2}>r_{M}$ and that $r_{M}>0$, we
conclude that the derivative $p'(r_{2})$ is strictly negative. In addition
to this, since $p(r)$ is a positive function that is the solution of a
first-order differential equation within $(r_{1},r_{2})$, it must be a
continuous and differentiable function. Therefore, given that it is zero
at both ends and always increases as we go to the interior of the
interval, it must have a point of maximum $r_{e}$ somewhere in the
interior of the interval, where we will have that $p'(r_{e})=0$. Using the
differential equation for $p(r)$ given by Equation~(\ref{Eqn38}) at this
point we thus obtain

\begin{equation}\label{Eqn53}
  \frac{1}{2r_{e}}
  \left[1+p(r_{e})\right]
  \frac
  {
    \Upsilon_{0}^{2}\left(r_{2}^{3}-r_{e}^{3}\right)-3r_{M}
    -
    3\Upsilon_{0}^{2}r_{e}^{3}p(r_{e})
  }
  {
    \Upsilon_{0}^{2}\left(r_{2}^{3}-r_{e}^{3}\right)
    +
    3\left(r_{e}-r_{M}\right)
  }
  =
  0.
\end{equation}

\noindent
This can only be zero if the numerator is zero, so we have that

\begin{equation}\label{Eqn54}
  \Upsilon_{0}^{2}r_{e}^{3}p(r_{e})
  =
  \frac{1}{3}\,
  \Upsilon_{0}^{2}
  \left(r_{2}^{3}-r_{e}^{3}\right)
  -
  r_{M}.
\end{equation}

\noindent
Since $\Upsilon_{0}^{2}>0$ and at its maximum we must have $p(r_{e})>0$
for the dimensionless pressure, we conclude that our lemma holds,

\begin{equation}\label{Eqn55}
  \frac{1}{3}\,
  \Upsilon_{0}^{2}
  \left(r_{2}^{3}-r_{e}^{3}\right)
  -
  r_{M}
  >
  0.
\end{equation}

\noindent
Let us now consider the result for $r_{\mu}$ in terms of the given
parameters of the problem, as shown in Equation~(\ref{Eqn26}), which we
can write as

\begin{equation}\label{Eqn56}
  r_{\mu}
  =
  \frac{1}{3}\,
  \Upsilon_{0}^{2}
  \left(r_{2}^{3}-r_{1}^{3}\right)
  -
  r_{M}.
\end{equation}

\noindent
By adding and subtracting terms to this equation, we can write it as

\begin{equation}\label{Eqn57}
  r_{\mu}
  =
  \left[
    \frac{1}{3}\,
    \Upsilon_{0}^{2}
    \left(r_{2}^{3}-r_{e}^{3}\right)
    -
    r_{M}
  \right]
  +
  \frac{1}{3}\,
  \Upsilon_{0}^{2}
  \left(r_{e}^{3}-r_{1}^{3}\right).
\end{equation}

\noindent
The quantity within square brackets is the one we just proved to be
strictly positive in our lemma. The other term is also strictly positive
because we certainly have that $r_{e}>r_{1}$. Therefore, we have our
theorem,

\begin{equation}\label{Eqn58}
  r_{\mu}
  >
  0.
\end{equation}

\noindent
Therefore, every solution of the problem that exists at all is bound to
have a singularity at the origin, which is characterized by the factor

\begin{equation}\label{Eqn59}
  \ln\!\left(\frac{r+r_{\mu}}{r}\right),
\end{equation}

\noindent
that appears with a negative sign in $\lambda_{i}(r)$ and with a positive
sign in $\nu_{i}(r)$. This implies that at this singular point we have
that

\noindent
\begin{eqnarray}\label{Eqn60}
  \lim_{r\to 0}\lambda_{i}(r)
  & = &
        -\infty,
        \nonumber
  \\
  \lim_{r\to 0}\e{\lambda_{i}(r)}
  & = &
        0,
        \nonumber
  \\
  \lim_{r\to 0}\nu_{i}(r)
  & = &
        \infty,
        \nonumber
  \\
  \lim_{r\to 0}\e{\nu_{i}(r)}
  & = &
        \infty.
\end{eqnarray}

\noindent
Note that this singularity does not have any disastrous consequences,
since it does not imply infinite concentrations of matter. In fact, we
have $\rho(r)=0$ in the whole inner vacuum region, including at the
origin. For the proper lengths in the radial direction, it just implies
that they get progressively more {\em contracted\/} as we approach the
origin, rather than being expanded with respect to the corresponding
variations of the radial coordinate $r$, as is the case in the outer
vacuum region. For the proper times it just means that we get
progressively more severe {\em red\/} shifts as we approach the origin,
rather than the blue shifts that we get as we approach the event horizon
from the outer vacuum region.

As a corollary to the proof that $r_{\mu}>0$, note that this fact
guarantees the positivity of the cubic polynomial in
Equation~(\ref{Eqn28}). This is so because the second derivative of that
polynomial is given by $-6\kappa\rho_{0}r$, being therefore negative for
all $r\in[r_{1},r_{2}]$. This means that the graph of the cubic polynomial
has a concavity turned downward throughout this interval. In addition to
this, it is easy to see that at $r=r_{2}$ the polynomial is given by
$3\left(r_{2}-r_{M}\right)$, which is strictly positive so long as
$r_{2}>r_{M}$. Finally, at $r=r_{1}$ the polynomial is given by

\begin{equation}\label{Eqn61}
  \kappa\rho_{0}\left(r_{2}^{3}-r_{1}^{3}\right)
  +
  3\left(r_{1}-r_{M}\right)
  =
  3\left(r_{1}+r_{\mu}\right),
\end{equation}

\noindent
where we used Equation~(\ref{Eqn26}), which is also strictly positive
since $r_{\mu}>0$. As a consequence of this, we may conclude that, so long
as the conditions $r_{2}>r_{M}$ and $r_{\mu}>0$ hold, as they must for
physically sensible solutions, the polynomial is strictly positive for all
$r\in[r_{1},r_{2}]$.

\subsection{Nature of the Inner Gravitational Field}\label{SSec0302}

The physical interpretation of the function $\nu(r)$ is that the proper
time interval at the radial position $r$, between two events occurring at
the same spatial point, is given by $d\tau=\exp[\nu(r)]dt$, where $dt$ is
the time interval between the two events as seen at spatial infinity,
where spacetime is flat. If we consider a photon traveling in the radial
direction, either inwards or outwards, this means that the proper
frequency $f(r)$ of the photon changes with position, between a first
point $r_{a}$ and a second point $r_{b}$, according to

\noindent
\begin{eqnarray}\label{Eqn62}
  f(r_{a})
  & = &
        \e{-\nu(r_{a})}f_{\infty},
        \nonumber
  \\
  f(r_{b})
  & = &
        \e{-\nu(r_{b})}f_{\infty},
\end{eqnarray}

\noindent
where $f_{\infty}$ is the frequency of the photon at radial infinity.
Dividing these two equations and making the two points very close
together, so that $r_{a}=r$ and $r_{b}=r_{a}+\delta r$, we have

\begin{equation}\label{Eqn63}
  \frac{f(r+\delta r)}{f(r)}
  =
  \e{-[\nu(r+\delta r)-\nu(r)]}.
\end{equation}

\noindent
For sufficiently small $\delta r$ we may write the variation of the
function $\nu(r)$ in terms of its derivative $\nu'(r)$, so that we get

\begin{equation}\label{Eqn64}
  \frac{f(r+\delta r)}{f(r)}
  \simeq
  \e{-\delta r\,\nu'(r)}.
\end{equation}

\noindent
Since the energy $hf(r)$ of a photon, $h$ being the Planck constant, is
proportional to its frequency, we have an interpretation of the red and
blue shifts of the frequency of the photons as decreases or increases in
their energies, respectively. We thus observe that, if a photon is going
outward, so that $\delta r>0$, and if the derivative $\nu'(r)$ is
positive, then we will have that $f(r+\delta r)<f(r)$, and therefore a red
shift in the frequency. If it is going outward but the derivative is
negative, then we will have that $f(r+\delta r)>f(r)$ and hence a blue
shift. On the other hand, if the photon is going inward, so that
$\delta r<0$, and the derivative is positive, then we will have a blue
shift, and finally, if it is going inward and the derivative is negative,
then we will have a red shift. Let us write down the derivative of
$\nu(r)$ in the inner and outer vacuum regions,

\noindent
\begin{eqnarray}\label{Eqn65}
  \nu'(r)
  & = &
        \left\{
        \begin{array}{rcl}
          -\,
          \FFrac{1}{2}\,
          \FFrac{r_{\mu}}{r(r+r_{\mu})}
          &
            \mbox{for}
          &
            0\;\leq r\leq r_{1},
          \\[3ex]
          \FFrac{1}{2}\,
          \FFrac{r_{M}}{r(r-r_{M})}
          &
            \mbox{for}
          &
            r_{2}\leq r<\infty.
        \end{array}
            \right.
\end{eqnarray}

\noindent
Let us now consider the consequences of Equation~(\ref{Eqn64}) in more
detail in each one of these two regions, starting with the outer vacuum
region. As one can see above, in the outer vacuum region, since we have
that $r>r_{2}>r_{M}>0$, the derivative $\nu'(r)$ is always positive.
Therefore, photons traveling outward undergo red shifts, while those
traveling inward undergo blue shifts. This can be interpreted in energetic
terms as the statement that when traveling inward the photons gain energy
from the gravitational field, and when traveling outward they lose energy
to it. This is characteristic of a gravitational field that is attractive
towards the origin.

However, in the inner vacuum region the situation is reversed. Since we
have that $r_{\mu}>0$, the derivative is everywhere {\em negative} in that
region. This means that photons traveling outward within this region are
{\em blue} shifted, and therefore {\em gain} energy from the gravitational
field, while photons traveling inward within this region are {\em red}
shifted, and therefore {\em lose} energy to the gravitational field. This
is characteristic of a gravitational field that is repulsive, driving
matter and energy away from the origin. This is the exact opposite of what
happens in the outer vacuum region. It is important to note that this
repulsion is not from the matter in itself, but from the {\em origin},
consisting therefore of an outward {\em attraction} towards the shell of
matter.

\section{Examples of Specific Solutions}\label{Sec04}

In order to calculate $z(r)$ either analytically or numerically it is
convenient to define a dimensionless variable $x$ such that

\noindent
\begin{eqnarray}\label{Eqn66}
  x
  & \equiv &
             \Upsilon_{0}\,r
             \;\;\;\Longrightarrow
             \nonumber
  \\
  \frac{d}{dr}
  & = &
        \Upsilon_{0}\frac{d}{dx}.
\end{eqnarray}

\noindent
In terms of $x$, Equation~(\ref{Eqn40}), that determines $z(r)$, becomes

\begin{equation}\label{Eqn67}
  z'(x)+\frac{\eta+2x^{3}}{2x(\eta+3x-x^3)}z(x)
  =
  \frac{3x^{3}}{2x(\eta+3x-x^3)},
\end{equation}

\noindent
where the primes indicate now derivatives with respect to $x$, and where
we define

\begin{equation}\label{Eqn68}
  \eta
  \equiv
  x_{2}^3-3 x_{M},
  \;\;
  x_{1}
  \equiv
  \Upsilon_{0}\,r_{1},
  \;\;
  x_{2}
  \equiv
  \Upsilon_{0}\,r_{2},
  \;\;
  x_{M}
  \equiv
  \Upsilon_{0}\,r_{M}.
\end{equation}

\noindent
Thus $x_{1}$, $x_{2}$ and $x_{M}$ correspond respectively to the internal
radius $r_{1}$, the external radius $r_{2}$ and the Schwarzschild radius
$r_{M}$, expressed in terms of the new variable $x$. The solution of
Equation~(\ref{Eqn67}) is obtained by writing Equation~(\ref{Eqn46}) in
terms of $x$,

\begin{equation}\label{Eqn69}
  z(x)
  =
  \sqrt{\frac{\eta+3x-x^3}{x}}
  \left[
    \sqrt{\frac{x_{2}}{3(x_{2}-x_{M})}}
    +
    \frac{3}{2}\,
    \int_{x_{2}}^{x}dy\,
    \frac{y^{5/2}}{(\eta+3y-y^{3})^{3/2}}
  \right],
\end{equation}

\noindent
where, in order to remain within the matter region, we must have
$x_{1}\le x\le x_{2}$. If we multiply both the numerator and the
denominator of the integral in Equation~(\ref{Eqn69}) by $y^{3/2}$, define
the polynomial $Q(y)=y\left(\eta+3y-y^{3}\right)$ and the rational
function $S(y,Q)\equiv y^{4}/Q^{3}$, then the integral in
Equation~(\ref{Eqn69}) can be rewritten as

\begin{equation}\label{Eqn70}
  \int_{x_{2}}^{x}dy\,
  \frac{y^{5/2}}{(\eta+3y-y^{3})^{3/2}}
  =
  \int_{x_{2}}^{x}
  S\!\left[y,\sqrt{Q(y)}\,\right]dy.
\end{equation}

\noindent
The expression on the right-hand side of Equation~(\ref{Eqn70}) is by
definition an elliptic integral~\cite{AbramowitzStegun} and cannot be
expressed in terms of elementary functions except in two cases: 1)
$S\!\left(y,Q^{1/2}\right)$ contains no odd powers of $y$; in our case
this happens when $\eta=0$ and leads to the Schwarzschild interior
solution; 2) the polynomial $Q(y)$ has two equal roots; this leads to the
explicit solutions that we discuss next.

\subsection{A Family of Explicit Solutions}\label{SSec0401}

The integral in Equation~(\ref{Eqn69}) contains a cubic polynomial. The
nature of its three roots depends on the value of its discriminant
$\Delta$~\cite{WikiCubicEq}. For cubic polynomials of the form $ax^3+cx+d$
we have $\Delta=-4ac^3-27a^{2}d^{2}$. If $\Delta>0$ the polynomial has
three distinct real roots, if $\Delta=0$ it has three real roots but two
of them are equal, and if $\Delta<0$ it has one real and two complex roots
which are conjugate to each other. In our case we have $a=-1$, $c=3$,
$d=\eta$ and therefore $\Delta=27(4-\eta^{2})$.

\begin{figure}[t]
  \centering
  {\color{white}\rule{\textwidth}{0.1ex}}
  % \fbox{
  % 
  \epsfig{file=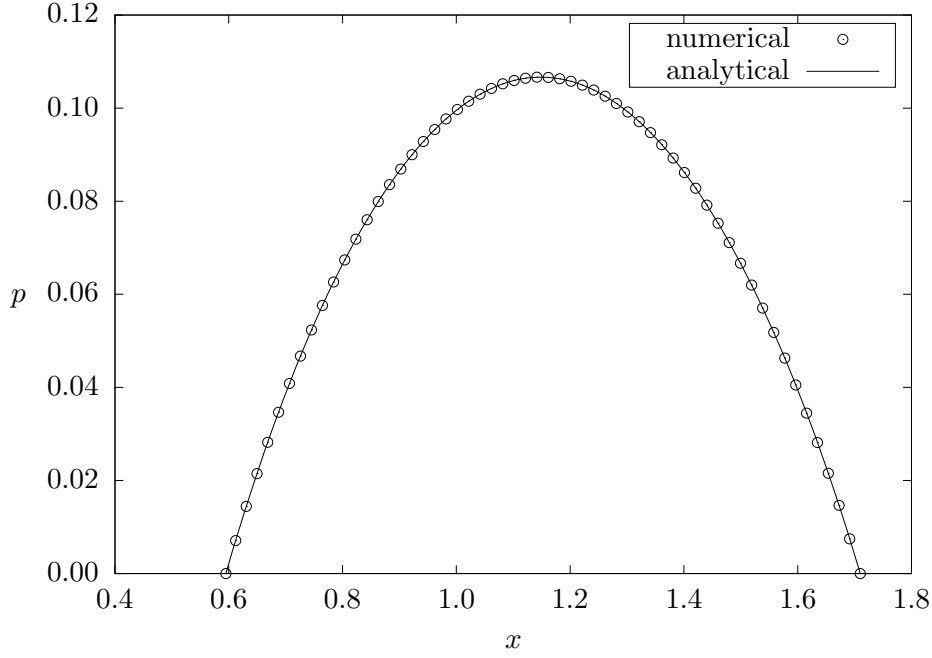,scale=1.0,angle=0}
  % 
  % }
  \caption{Comparison between the dimensionless pressure $p(x)$ calculated
    analytically and numerically using the Runge-Kutta fourth-order
    algorithm for $\eta=2.0$, $x_{2}=5^{1/3}$, $x_{M}=1.0$ and
    $x_{1}=0.594881$.}
  \label{Fig01}
\end{figure}

\begin{figure}[t]
  \centering
  {\color{white}\rule{\textwidth}{0.1ex}}
  % \fbox{
  % 
  \epsfig{file=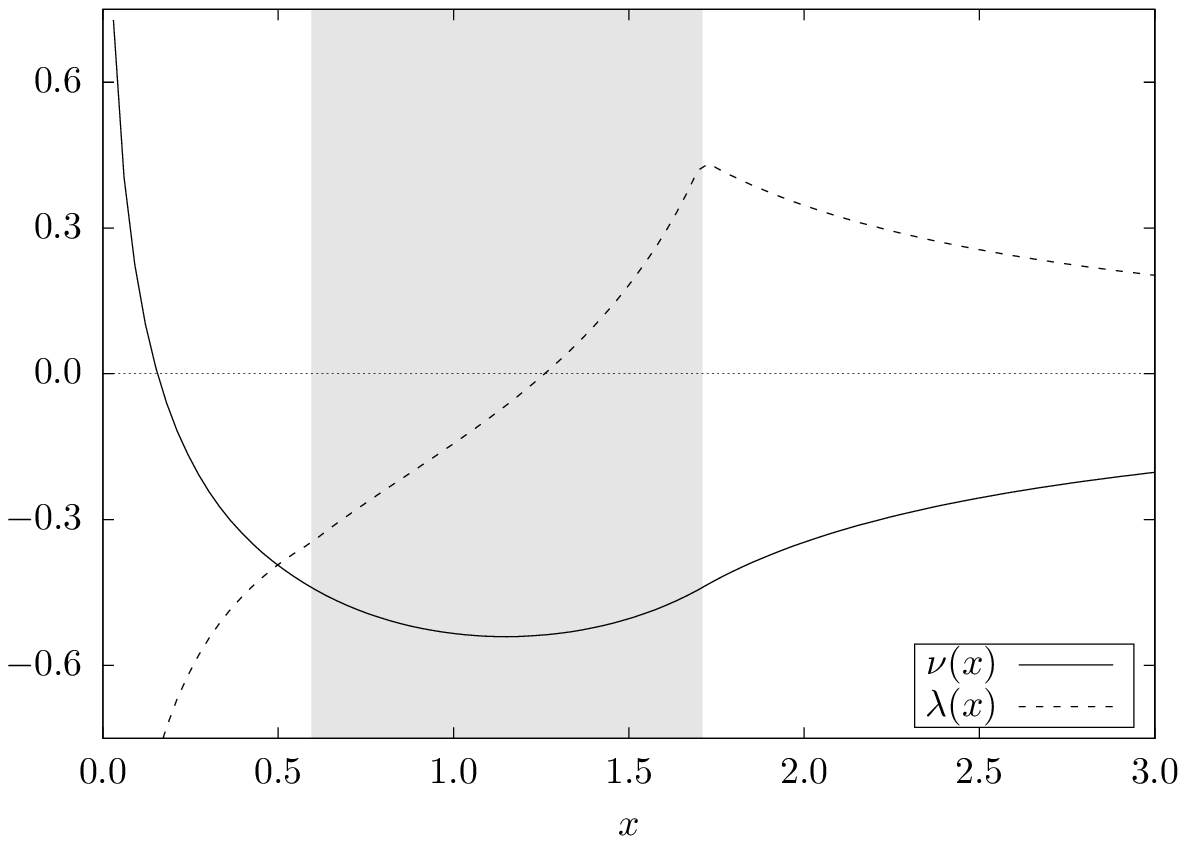,scale=1.0,angle=0}
  % 
  % }
  \caption{The functions $\nu(x)$ and $\lambda(x)$ for $\eta=2.0$,
    $x_{2}=5^{1/3}$ and $x_{M}=1.0$. The shaded area indicates the matter
    region, to its right is the outer vacuum and to its left is the inner
    vacuum.}
  \label{Fig02}
\end{figure}

\begin{figure}[t]
  \centering
  {\color{white}\rule{\textwidth}{0.1ex}}
  % \fbox{
  % 
  \epsfig{file=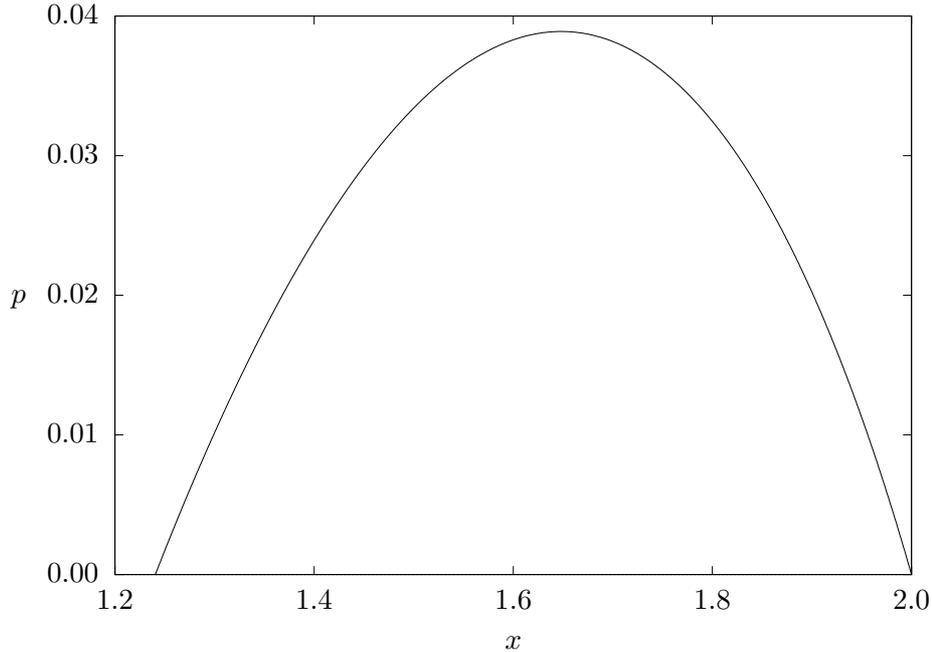,scale=1.0,angle=0}
  % 
  % }
  \caption[]{The dimensionless pressure $p$ calculated numerically for
    $\eta=5.0$, $x_{2}=2.0$ and $x_{M}=1.0$.}
  \label{Fig03}
\end{figure}

\begin{figure}[t]
  \centering
  {\color{white}\rule{\textwidth}{0.1ex}}
  % \fbox{
  % 
  \epsfig{file=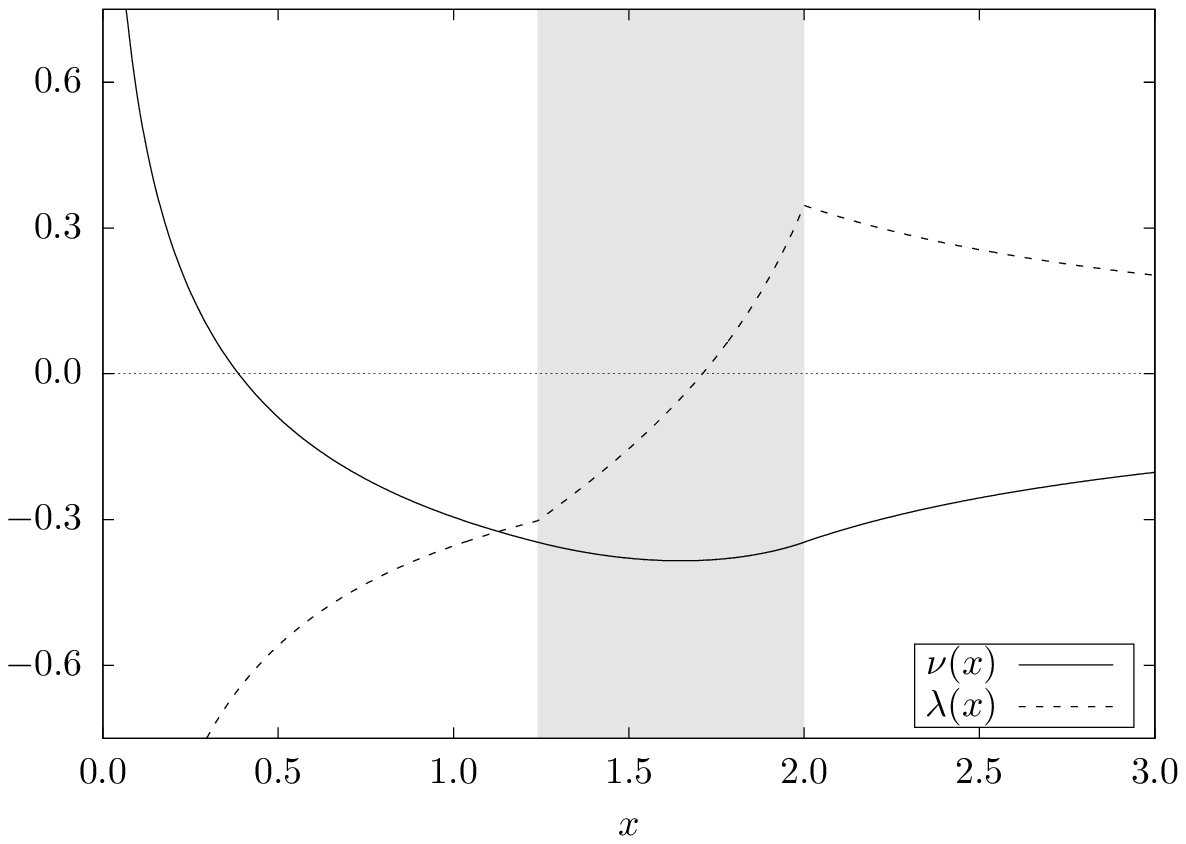,scale=1.0,angle=0}
  % 
  % }
  \caption[]{The functions $\nu(x)$ and $\lambda(x)$ for $\eta=5.0$,
    $x_{2}=2.0$ and $x_{M}=1.0$. The shaded area indicates the matter
    region, to its right is the outer vacuum and to its left is the inner
    vacuum.}
  \label{Fig04}
\end{figure}

The value $\Delta=0$ corresponds to the case where the solution for $z(x)$
can be expressed in terms of elementary functions. Note that we have
$\Delta=0$ when $\eta=\pm 2$, which corresponds to
$x_{2}^3=\pm2+3x_{M}$. For $\eta=-2$ the polynomial in the integral of
Equation~(\ref{Eqn69}) is non-positive for $x\ge 0$. Therefore, we must
choose $\eta=2$. For this value of $\eta$ the polynomial is strictly
positive in the interval $[0,2)$ and can be factored as

\begin{equation}\label{Eqn71}
  2+3y-y^{3}
  =
  (2-y)(y+1)^{2}.
\end{equation}

\noindent
In this case we can express the integral in Equation~(\ref{Eqn69}) in
terms of elementary functions. The calculation can be considerably
simplified using a new integration variable $u$ defined by
$u=\sqrt{y/(2-y)}$. The final result, up to an integration constant, is

\noindent
\begin{eqnarray}\label{Eqn72}
  \mathcal{I}(y)
  & \equiv &
             \int\,dy\,
             \frac{y^{5/2}}{(2-y)^{3/2}(y+1)^3}
             \nonumber
  \\
  & = &
        \frac{2y^{2}+15y+10}{18\,(y+1)^{2}}\sqrt{\frac{y}{2-y}}
        -\,
        \frac{5\sqrt{3}}{27}
        \arctan\!\left(\sqrt{\frac{3y}{2-y}}\,\right).
\end{eqnarray}

\noindent
Thus, in terms of $\mathcal{I}(y)$ Equation~(\ref{Eqn69}) reads

\begin{equation}\label{Eqn73}
  z(x)
  =
  \sqrt{\frac{2+3x-x^3}{x}}
  \left\{
    \sqrt{\frac{x_{2}}{3(x_{2}-x_{M})}}
    +
    \frac{3}{2}
    \left[\rule{0em}{2.5ex}\mathcal{I}(x)-\mathcal{I}(x_{2})\right]
  \right\}.
\end{equation}

\noindent
Note that, in order to guarantee that the cubic polynomial for $\eta=2$
shown in Equation~(\ref{Eqn71}) is always positive, we need to have $y<2$.
Therefore, since we already know that the polynomial is positive, the
arguments of the square roots in Equation~(\ref{Eqn72}) are always
positive.

\subsection{Examples of Numerical Solutions}\label{SSec0402}

In our numerical approach here, we assume that the external radius
$x_{2}=\Upsilon_{0}r_{2}$ is given. In order to complete the calculation
we have to determine the interior radius $x_{1}$. This can be done
recalling that the dimensionless pressure $p(x)$ is zero for $x=x_{2}$ and
$x=x_{1}$. Since according to Equation~(\ref{Eqn47}) $p(x)=1/z(x)-1$, this
is equivalent to the determination of the values of $x$ for which
$z(x)=1$. By the determination of $x_{1}$ we would have solved the problem
in the entire matter region. Note that since
$x=\Upsilon_{0}r=\sqrt{\kappa\rho_{0}}\;r$ we have obtained a family of
solutions parametrized by two parameters, the external radius $r_{2}$ and
the parameter $\eta$.

If the discriminant $\Delta\ne 0$ the integral in Equation~(\ref{Eqn69})
is expressed in terms of elliptic integrals and the result is not very
transparent. It is more convenient to integrate the differential
Equation~(\ref{Eqn67}) using the fourth-order Runge-Kutta algorithm
(RK4)~\cite{NumericalRecipes}. We start by choosing a value of $x=x_{2}$
for which the cubic polynomial is positive and we put $z(x_{2})=1$. This
determines the outer radius of the matter shell. We then iterate the
differential equation given in Equation~(\ref{Eqn67}) in the decreasing
$x$ direction until we reach the first point for which the value of $z$
returns to $1$. This point is chosen as $x_{1}$. If a value for $x_{1}$
cannot be found, we conclude that there is no solution to the problem with
the given values of $x_{2}$ and $x_{M}$. A good test for the efficiency of
the algorithm is to compare the exact analytic result given in
Equation~(\ref{Eqn73}) with the result from the numerical integration in
that same case. These results are shown in Figure~\ref{Fig01}. On any
current $64$-bit desktop computer one can easily reach a high degree of
precision with little numerical effort. After iterating the RK4 algorithm
from $x_{2}$ to $x_{1}$ the difference between the exact and the numerical
results for $z(x)$ stays below $1.03536\times 10^{-29}$ for an iteration
step of $\delta x\approx 10^{-7}$.

In the comments that follow $x_{\mu}\equiv\Upsilon_{0}r_\mu$, where
$r_\mu$ is the integration constant that results from the solution of the
Einstein equations in the inner vacuum region, given in
Equation~(\ref{Eqn26}). In the matter region the input parameters are
$\eta$ and $x_{2}$. The parameter $x_{1}$ is obtained from the iteration
of Equation~(\ref{Eqn67}). The value of $x_{M}$ that is necessary for
plotting the curves is given in Equation~(\ref{Eqn68}). The expressions
for $\lambda(x)$ and $\nu(x)$ are given in Table~\ref{Tab01}.
Figure~\ref{Fig02} shows the plots of the functions $\nu(x)$ and
$\lambda(x)$ for $\eta=2.0$ and $x_{2}=5^{1/3}$. The curves were obtained
analytically using Equation~(\ref{Eqn73}) and the expressions in
Table~\ref{Tab01}, but using the numerically calculated parameters
$x_{1}=0.594881$ and $x_{\mu}=0.596494$.

In Figure~\ref{Fig03} we plot the dimensionless pressure $p(x)$ as a
function of $x$, in a case in which there is no analytic expression and
the calculation is performed numerically. The parameters are
$x_{1}=1.24050$ and $x_{\mu}=1.03035$. Comparing Figures~\ref{Fig01}
and~\ref{Fig03}, that depict the dimensionless pressure $p(x)$ as a
function of $x$ for $\eta=2.0$ and $\eta=5.0$, one notes that the two
graphs are similar but for larger values of $\eta$ the graph becomes less
symmetric.

Figure~\ref{Fig04} shows the plots of the functions $\nu(x)$ and
$\lambda(x)$, for $\eta=5.0$ and $x_{2}=2.0$. In this case there are no
analytical solutions available in the matter region and the values of
$\nu(x)$ and $\lambda(x)$ were obtained numerically. In the vacuum regions
we used the analytical expressions given in Table~\ref{Tab01} with the
parameters $x_{1}=1.24050$ and $x_{\mu}=1.03035$.

\section{Conclusions}\label{Sec05}

In this paper we have given the complete and exact solution of the
Einstein field equations for the case of a shell of liquid matter.
Although this particular problem can be seen as having a somewhat academic
nature, it does lead us to two important and unexpected conclusions. One
of them is that all solutions for shells of liquid matter have a
singularity at the origin, within the inner vacuum region, that does {\em
  not\/}, however, lead to any kind of pathological behavior involving the
matter. The other is that, contrary to what is usually thought, a
non-trivial gravitational field does exist within a spherically symmetric
central cavity, namely the inner vacuum region.

The geometry within the cavity is associated with a spacetime that is
contracted in the radial direction, rather than expanded. It is easy to
verify that, unlike what happens in the outer vacuum region, the proper
radial length, $\ell_{1}$, say from $r=0$ to $r=r_{1}$, is in fact {\em
  smaller} than the corresponding radial coordinate $r_{1}$. We have that
$d\ell_{1}=\sqrt{g_{11}}\,dr$, and therefore

\noindent
\begin{eqnarray}\label{Eqn74}
  \ell_{1}
  & = &
        \int_{0}^{r_{1}}dr\,
        \sqrt{\frac{r}{r+r_{\mu}}}
        \nonumber
  \\
  & < &
        \int_{0}^{r_{1}}dr
        \nonumber
  \\
  & = &
        r_{1},
\end{eqnarray}

\noindent
given that $r_{\mu}>0$. This illustrates the fact that the radial lengths
within the inner vacuum region are contracted rather than expanded. The
true physical volume of the inner vacuum region is therefore
correspondingly smaller than the apparent coordinate volume. This renders
this inner geometry not embeddable in the illustrative way that is usually
employed in the case of the outer vacuum region.

The gravitational field associated to this geometry, inside the inner
vacuum region, can be interpreted as a repulsive field with respect to the
origin. This can be ascertained from an examination of the sign of the
derivative of $\nu(r)$ in the inner and outer vacuum regions, and its
interpretation in terms of the energy of a photon traveling in the radial
direction. This sign is positive in the outer vacuum region, corresponding
to an attractive field towards the origin, and negative in the inner
vacuum region, corresponding to an repulsive field away from the origin.
Of course, since $\nu'(r)$ is a continuous function, and since we enter
the matter region from the outer vacuum region with a positive derivative,
and exit it into the inner vacuum region with a negative derivative, there
must be a point within the matter region where $\nu'(r)=0$, and where the
derivative flips sign. This is clearly the point $r_{e}$ of minimum of
$\nu(r)$, which is also the point of minimum of $z(r)$, and hence the
point of maximum of the pressure $P(r)$, a point which already had a role
to play in our arguments.

One can acquire an intuitive understanding of the unexpected situation in
the inner vacuum region by observing that such a situation can arise even
within the Newtonian framework, if we use a slightly modified potential.
We can do this if we consider the Newtonian argument for the gravitational
force within a hollow thin spherical shell of matter, but with a potential
that behaves as $1/r^{1+\epsilon}$ for some $|\epsilon|\ll 1$, thus
leading to a force that behaves as $1/r^{2+\epsilon}$. If one considers a
test mass at a point in the interior of the hollow shell, at the position
$\vec{r}$ with respect to the center, it is not difficult to use the usual
Newtonian argument to show that, if $\epsilon>0$, then the resulting
gravitational force at that point is oriented outward, in the direction of
$\vec{r}$, towards the shell of matter. In other words, the attraction by
the part of the shell that is closer to the point $\vec{r}$ outweighs the
attraction from the opposite side, thus leading to a resulting force that
repels particles away from the origin. Note that this argument involving a
potential behaving in a way other than $1/r$ is the same that can be used
to model the precession of the perihelion of orbits in General Relativity
using this semiclassical Newtonian framework. That precession is prograde
precisely if $\epsilon>0$.

It is interesting to note that this configuration of the gravitational
field tends to stabilize the shell of liquid matter, since any particle of
matter that detaches from the liquid and wanders into one of the vacuum
regions will be driven back to the bulk of the liquid by the gravitational
field. This can be interpreted as a successful stability test satisfied by
all the solutions. The general tendency of the gravitational field is
therefore that of compressing the shell of fluid matter, from both sides.
This suggests that the same interpretation should be valid in the case of
a gaseous fluid.

The singularity at the origin is usually thought to be associated with an
infinite concentration of matter there, and thus considered to be an evil
that must be avoided at any cost. However, this only makes any sense at
all if one thinks of that singularity as a point of gravitational
attraction, rather than as a point of repulsion of matter. Here we do have
the singularity, but not the infinite concentration of matter at the
origin, due to the repulsive character of the gravitational field around
the origin. In any case, the existence of the singularity is not a
question of choice, of course, since it is required by the field equations
and by the interface boundary conditions that follow from them. One is not
at liberty to impose that $r_{\mu}=0$ in order to avoid this singularity.

\bibliography{allrefs_en}\bibliographystyle{ieeetr}

\end{document}